\documentclass[prl,final,twocolumn,showpacs,floatfix]{revtex4}%
\usepackage{amsfonts}
\usepackage{amsmath}
\usepackage{amssymb}
\usepackage{graphicx}%
\providecommand{\U}[1]{\protect\rule{.1in}{.1in}}

\def \nobreakseq {\nobreak \hskip 0pt \hbox}

\begin{document}
\title{Structure and rotations of the Hoyle state}
\author{Evgeny Epelbaum$^{a}$, Hermann~Krebs$^{a}$, Timo L\"{a}hde$^{b}$,
Dean~Lee$^{d}$, Ulf-G.~Mei{\ss }ner$^{e,b,c}$}
\affiliation{$^{a}$%
Institut~f\"{u}r~Theoretische~Physik~II,~Ruhr-Universit\"{a}t~Bochum,~D-44870~Bochum,~Germany\linebreak%
$^{b}$Institut~f\"{u}r~\nobreakseq{Kernphysik,~Institute}~for~Advanced~Simulation,\linebreak
J\"{u}lich~Center~for~Hadron~Physics,~Forschungszentrum~J\"{u}lich,~D-52425~J\"{u}lich,~Germany\linebreak
$^{c}$JARA~-~High~Performance~Computing,~Forschungszentrum~J\"{u}lich,~D-52425~J\"{u}lich,~Germany\linebreak
$^{d}$Department~of~Physics,~North~Carolina~State~University,~Raleigh,~NC~27695,~USA\linebreak%
$^{e}$%
Helmholtz-Institut~f\"{u}r~Strahlen-~und~Kernphysik~and~Bethe~Center~for~Theoretical~Physics,\linebreak%
~Universit\"{a}t~Bonn,~D-53115~Bonn,~Germany\linebreak}

\begin{abstract}
The excited state of the $^{12}$C nucleus known as the ``Hoyle state'' constitutes one of the most interesting, 
difficult and timely challenges in nuclear physics, as it plays a key role in the production of carbon via fusion
of three alpha particles in red giant stars. In this letter, we present \textit{ab initio} lattice calculations which 
unravel the structure of the Hoyle state, along with evidence for a low-lying \mbox{spin-2} rotational excitation. 
For the $^{12}$C ground state and the first excited \mbox{spin-2} state, we find a compact triangular configuration 
of alpha clusters. For the Hoyle state and the second excited \mbox{spin-2} state, we find a ``bent-arm'' or
obtuse triangular configuration of alpha clusters. We also calculate the electromagnetic transition rates 
between the low-lying states of $^{12}$C.
\end{abstract}

\maketitle

The carbon nucleus $^{12}$C is produced by fusion of three alpha particles
in red giant stars. However,
without resonant enhancement the triple alpha reaction is too slow to account for
the observed abundance of carbon in the Universe. In the early 1950's, \"{O}pik and Salpeter noted 
independently that the first step of merging two alpha particles is enhanced by the
formation of $^{8}$Be \cite{Opik:1951,Salpeter:1952,Salpeter:1953}. The
ground state of $^{8}$Be is a resonance with energy $92$~keV above the $^{4}%
$He-$^{4}$He threshold and a width of $2.5$~eV. However, a year later Hoyle 
realized that this enhancement is still insufficient. 
To resolve this discrepancy, Hoyle predicted 
an unobserved positive-parity resonance of $^{12}$C just above the combined masses 
of $^{8}$Be and $^{4}$He~\cite{Hoyle:1954zz}.

About three years later, Cook, Fowler, Lauritsen and Lauritsen
observed a $J^{\pi}=0^{+}$ state $278$~keV above the $^{8}%
$Be-$^{4}$He threshold~\cite{Cook:1957}. This excited $0_{2}^{+}$ state has
a width of $8.5$~eV, and is now commonly known as the ``Hoyle state''. The
triple alpha reaction is completed when the Hoyle state decays
electromagnetically to the $2_{1}^{+}$ state and subsequently to the
$0_{1}^{+}$ ground state. Around the same time, Morinaga conjectured that the structure of excited
alpha-nuclei such as the Hoyle state may be non-spherical~\cite{Morinaga:1956},
which would imply low-lying rotational excitations of even parity.
Other ideas also exist for the structure of the Hoyle state, such as a diffuse trimer of 
alpha particles~\cite{Tohsaki:2001an}. Recently, the spin-2 excitation of the 
Hoyle state has attracted considerable interest from several experimental 
groups~\cite{Freer:2009,Hyldegaard:2010,Zimmerman:2011,Itoh:2011,Smit:2012zk}.

We have recently presented an \textit{ab initio} lattice calculation of the
Hoyle state~\cite{Epelbaum:2011md} where the low-lying spectrum of
$^{12}$C was explored using the framework of chiral effective field theory
and Monte Carlo lattice calculations.  However the central question regarding 
the alpha cluster structure of the Hoyle state remained unsolved, perhaps the 
greatest remaining challenge in \textit{ab initio} nuclear theory.  In this letter, 
we announce a major innovation of the lattice method that constructs and tests a 
wide class of nuclear wave functions explicitly.  We present 
\textit{ab initio} lattice results that resolve questions about the structure 
of the Hoyle state and the existence of
rotational excitations.  We also find evidence for a low-lying \mbox{spin-2} rotational
excitation of the Hoyle state. For the Hoyle state and its \mbox{spin-2}
excitation, we find strong overlap with a ``bent-arm'' or obtuse triangular
configuration of alpha clusters. This is in contrast with the $^{12}$C
ground state and the first \mbox{spin-2} state, where we note strong overlap with a
compact triangular configuration of alpha clusters. We also calculate the
electromagnetic transition rates among the low-lying even-parity states of
$^{12}$C. Our lattice results can be compared with other recent theoretical
calculations for the low-lying spectrum of $^{12}$C using the no-core shell
model~\cite{Roth:2011ar,Forssen:2011dr} and variational calculations using
Fermionic Molecular Dynamics~\cite{Chernykh:2007a,Chernykh:2010zu}.

Chiral effective field theory treats the interactions of protons and neutrons
as a systematic expansion in powers of nucleon momenta and the pion mass. A
recent review can be found in Ref.~\cite{Epelbaum:2008ga}. The low-energy
expansion is organized in powers of $Q$, where $Q$ denotes the typical
particle momentum and is treated on the same footing as the mass of the
pion. The most important contributions to the nuclear Hamiltonian appear at
leading order (LO) or $\mathcal{O}(Q^{0})$, while the next-to-leading order (NLO) terms
are $\mathcal{O}(Q^{2})$. In the lattice calculations presented here, we include all
possible interactions up to next-to-next-to-leading order (NNLO), or 
$\mathcal{O}(Q^{3})$.

Our analysis makes use of a periodic cubic lattice with a lattice spacing
of $a=1.97$~fm and a length of $L=12$~fm. In the time direction, we use
a step size of $a_{t}^{}=1.32~$fm and vary the propagation time $L_{t}^{}$ to
extrapolate to the limit $L_{t} \to \infty$. The nucleons are treated
as point-like particles on lattice sites, and interactions due to the exchange
of pions and multi-nucleon operators are generated using auxiliary 
fields. Lattice effective field theory was originally used to calculate the
many-body properties of homogeneous nuclear and neutron 
matter~\cite{Muller:1999cp,Lee:2004si}. Since then, the properties of several 
atomic nuclei have been investigated~\cite{Epelbaum:2009zs,Epelbaum:2009pd}. 
A recent review of the literature can be found in Ref.~\cite{Lee:2008fa}.

We use Euclidean time propagation to project on to low-energy states of the
interacting system. Let $H$ be the Hamiltonian. For any initial quantum
state $\Psi$, the projection amplitude is defined as the expectation value
$\left\langle e^{-Ht}\right\rangle_{\Psi}$.~For large Euclidean time $t$,
the exponential operator $e^{-Ht}$ enhances the signal of low-energy states.
Energies can then be determined from the exponential decay of these 
amplitudes. The first and last few time steps are evaluated
using a simpler Hamiltonian $H_{\text{SU(4)}}$, based upon the 
Wigner~SU(4) symmetry for protons and neutrons~\cite{Wigner:1937}. Such a
Hamiltonian is computationally inexpensive and is used as a low-energy filter
before the main calculation. This technique is described in Ref.~\cite{Lee:2008fa}.
\begin{table}[h]
\caption{Lattice results and experimental values for the ground state energies
of $^{4}$He and $^{8}$Be, in units of MeV. The quoted errors are one
standard deviation estimates which include both Monte Carlo statistical errors
and uncertainties due to extrapolation at large Euclidean time. 
\label{He4_Be8}
}
\vspace{.2cm}
\begin{tabular}{|c||c|c|} \hline
& \hspace{.4cm} $^{4}$He \hspace{.4cm} 
& \hspace{.4cm} $^{8}$Be \hspace{.4cm} \\ \hline
LO [$\mathcal{O}(Q^{0})$] & $-28.0(3)$ & $-57(2)$ \\
NLO [$\mathcal{O}(Q^{2})$] & $-24.9(5)$ & $-47(2)$ \\
NNLO [$\mathcal{O}(Q^{3})$] & $-28.3(6)$ & $-55(2)$ \\
Exp & $-28.30$ & $-56.50$ \\ \hline
\end{tabular}
\end{table}

In Table~\ref{He4_Be8}, we present lattice results for the ground state
energies of $^{4}$He and $^{8}$Be up to NNLO. The method of calculation is
nearly identical to that described in Refs.~\cite{Epelbaum:2009pd,Epelbaum:2010xt,Epelbaum:2011md}. As discussed in these references, our LO result already includes a large portion of the corrections usually counted at NLO.  Therefore the scaling of corrections with increasing order decreases much more rapidly than suggested by the results in Table~\ref{He4_Be8}.  
The higher-order corrections are computed using perturbation theory, and the
coefficients of the nucleon-nucleon interaction are determined by fitting available
low-energy scattering data. In our calculations the NNLO corrections
correspond to three-nucleon forces. A detailed description of the
interactions at each order can be found in Ref.~\cite{Epelbaum:2010xt}. We
have used the triton binding energy and the weak axial vector current to fix
the low-energy constants $c_{D}^{}$ and $c_{E}^{}$ which enter the three-nucleon
interaction.
In comparison with Ref.~\cite{Epelbaum:2011md}, improvements have been made 
using higher-derivative lattice operators which eliminate the overbinding of the LO action 
for larger nuclei such as $^{16}$O. The details of this improved action will be
discussed in a forthcoming publication. We note that
the computed binding energies for $^{4}$He and $^{8}$Be at NNLO are in agreement
with the experimental values.
\begin{figure}[t]
\includegraphics[trim=0.5cm 3cm 0.5cm 13cm,
width=8cm]{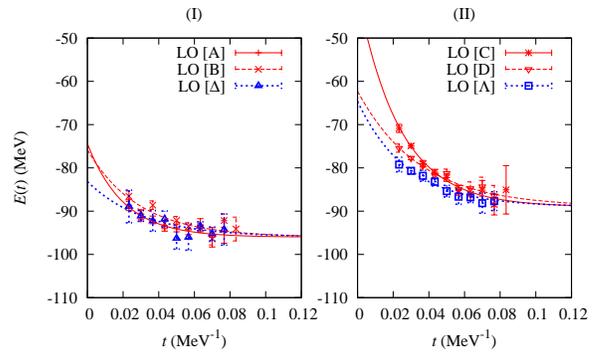}
\caption{Lattice results for the $^{12}$C spectrum at leading order~(LO). 
Panel~I shows the results using initial states $A$, $B$ and $\Delta$,
each of which approaches the ground state energy. Panel~II shows the results 
using initial states $C$, $D$ and $\Lambda$. These trace out an
intermediate plateau at an energy $\simeq 7$~MeV above the ground state.
\label{carbon12_0+_lo}}
\end{figure}

In our projection Monte Carlo calculations, we use a larger class of initial
and final states than hitherto considered. For $^{4}$He, we use an initial state 
with four nucleons, each at zero momentum. For $^{8}$Be, we use the same 
initial state as for $^{4}$He, followed by application of creation operators after the 
first Euclidean time step in order to inject four more nucleons at zero momentum.
An analogous procedure is performed in order to extract four nucleons before the last 
Euclidean time step. Such injection and extraction of nucleons at zero momentum helps 
to eliminate directional biases caused by the initial and final state momenta.

We make use of many different initial and final states in order to probe the structure
of the various $^{12}$C states. For the $^{12}$C states investigated here, we
measure four-nucleon correlations by calculating the expectation 
value $\langle \rho^{4} \rangle$, where $\rho$ is the total nucleon density.
We find strong four-nucleon correlations consistent with the formation of 
alpha clusters. In Fig.~\ref{carbon12_0+_lo}, we present lattice results for the 
energy of $^{12}$C at leading order versus Euclidean projection time~$t$. 
For each of the initial states $A-D$, we start with delocalized nucleon standing
waves and use a strong attractive interaction in $H_{\text{SU(4)}}$ to allow
the nucleons to self-organize into a nucleus. For the initial states $\Delta$
and $\Lambda$, we use alpha cluster wave functions to recover the same 
states found using the initial states $A-D$. For these calculations, the
interaction in $H_{\text{SU(4)}}$ is not as strong and the projected states
retain their original alpha cluster character.

In Panel~I of Fig.~\ref{carbon12_0+_lo}, we show lattice results corresponding
to the initial states $A$, $B$, and $\Delta$, each approaching the ground state 
energy $-96(2)$~MeV. For initial state~$A$, we start with four nucleons (each at 
zero momentum) apply creation operators after the first time step to inject four 
more nucleons at rest, followed by the injection of four additional nucleons at rest 
after the second time step. This procedure is used in reverse to extract nucleons 
for the final state~$A$. An identical scheme is used for initial state~$B$, though the 
interactions in $H_{\text{SU(4)}}$ are not as strongly attractive as those for~$A$.
\begin{figure}[b]
\centering{
\includegraphics[width=3cm]{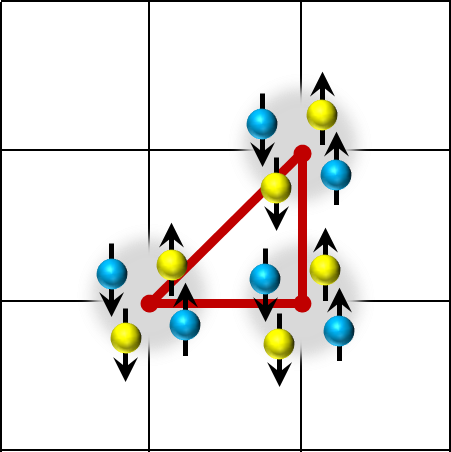}
\caption{Illustration of the initial state $\Delta$.  There are $12$ equivalent 
orientations of this compact triangular configuration. \label{ground}
}}
\end{figure}
\begin{figure}[t]
\centering{
\includegraphics[width=3cm]{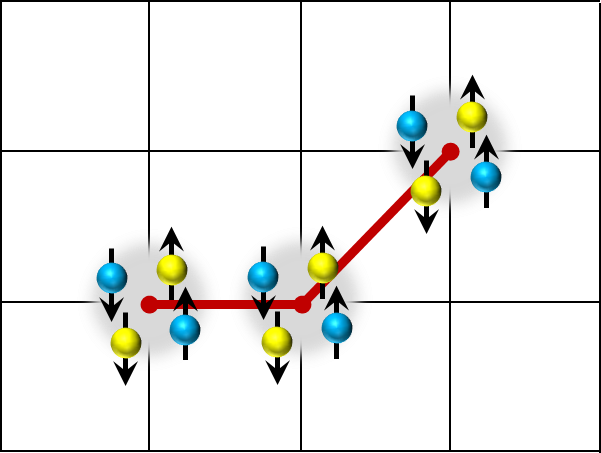}
\caption{Illustration of the initial state $\Lambda$.  There are $24$ equivalent 
orientations of this ``bent-arm'' or obtuse triangular configuration. \label{hoyle}
}}
\end{figure}

For the initial state $\Delta$, we use a wave function constructed from of three 
alpha clusters, as shown in Fig.~\ref{ground}. The alpha clusters are formed by
Gaussian wave packets centered on the vertices of a compact triangle. In order to
construct eigenstates of total momentum and lattice cubic rotations, we
consider all possible translations and rotations of the initial state. There
are a total of $12$ equivalent orientations of $\Delta$. We do not
find rapid convergence to the ground state when starting from any other
configuration of alpha clusters. We thus conclude that the alpha
cluster configurations in Fig.~\ref{ground} have the strongest overlap with
the $0_{1}^{+}$ ground state of $^{12}$C. The fact that $\Delta$ is an isosceles
right triangle rather than an equilateral triangle is merely a lattice artifact.

In Panel II of Fig.~\ref{carbon12_0+_lo}, we show the leading-order energies 
obtained from the initial states $C$, $D$, and $\Lambda$. Each of these
approaches an intermediate plateau at $-89(2)$~MeV. In the limit of infinite Euclidean 
time, these would eventually also approach the ground state energy. 
However, it is clear that a different state is first being
formed which is distinct from the ground state. We identify the $0^{+}$ state in this
plateau region as the $0_{2}^{+}$ Hoyle state. The common thread connecting
the initial states $C$, $D$, and $\Lambda$ is that each produces a state
which has an extended (or prolate) geometry. This is in contrast to the oblate
triangular configuration shown in Fig.~\ref{ground}.

For initial state~$C$, we take four nucleons at rest, four with momenta
$(2\pi/L,2\pi/L,2\pi/L)$, and four with momenta $(-2\pi/L,-2\pi/L,-2\pi/L)$.
Similarly, for initial state~$D$ we take four nucleons at rest, four
with momenta $(2\pi/L,2\pi/L,0)$, and four with momenta $(-2\pi/L,-2\pi/L,0)$.
Finally, initial state $\Lambda$ uses a set of three alpha clusters formed by
Gaussian wave packets centered on the vertices of a ``bent-arm'' (or obtuse) 
triangular configuration, as shown in Fig.~\ref{hoyle}. There are a total of $24$
equivalent orientations of $\Lambda$. We do not find the same energy
plateau starting from other configurations of alpha clusters. We conclude
that configurations of the type shown in Fig.~\ref{hoyle} have the strongest overlap with
the $0_{2}^{+}$ Hoyle state of $^{12}$C.
\begin{table}[h]
\caption{Lattice and experimental results for the energies of the low-lying even-parity states 
of $^{12}$C, in units of MeV. \label{c12_spectrum}}
\vspace{.2cm}
\begin{tabular}{|c||c|c|c|c|} \hline
& \hspace{.3cm} $0_{1}^{+}$ \hspace{.3cm}
& \hspace{.1cm} $2_{1}^{+}(E^{+})$ \hspace{.1cm} 
& \hspace{.3cm} $0_{2}^{+}$ \hspace{.3cm}
& $2_{2}^{+}(E^{+}) $ \\ \hline
LO 
& $-96(2)$ & $-94(2)$ & $-89(2)$ & $-88(2)$ \\
NLO 
& $-77(3)$ & $-74(3)$ & $-72(3)$ & $-70(3)$ \\
NNLO 
& $-92(3)$ & $-89(3)$ & $-85(3)$ & $-83(3)$ \\ \hline
Exp & $-92.16$ & $-87.72$ & $-84.51$ &
$\left.
\begin{tabular}{l}
$-82.6(1)$ \cite{Freer:2009,Zimmerman:2011} \\
$-81.1(3)$ \cite{Hyldegaard:2010} \\
$-82.32(6)$ \cite{Itoh:2011}
\end{tabular}
\right. $
\\ \hline
\end{tabular}
\end{table}

We use the multi-channel method of Ref.~\cite{Epelbaum:2011md}
to find a spin-2 excitation above the ground state, as well as a spin-2
excitation above the Hoyle state. In both cases, we make use of the $E^{+}$
representation of the cubic rotation group on the lattice. We summarize our
results for the binding energies of the low-lying, even-parity states of
$^{12}$C in Table~\ref{c12_spectrum}. The binding energies at
NNLO are in agreement with the experimental values.

In Table~\ref{radii}, we present results at LO for the
root-mean-square charge radii and quadrupole moments of the even-parity states
of $^{12}$C, with experimental values given where available. In this
study, we compute electromagnetic moments only at LO. We note that
moments such as the charge radius for resonances above threshold are dependent
on the boundary conditions used to regulate the continuum-state asymptotics of the
wave function. We avoid this problem, because all of the low-lying states are
bound at LO. One expects that as the higher-order corrections
push the binding energies closer to the triple alpha threshold, the
corresponding radii will increase accordingly. A detailed study of these
resonances as a function of lattice volume will be undertaken in future
work. We find good agreement with the experimental value for the $2_{1}^{+}$
quadrupole moment. The sign difference of the electric quadrupole moments
of the spin-2 states is a reflection of the oblate shape of the $2_{1}^{+}$ state
and the prolate shape of the $2_{2}^{+}$ state.
\begin{table}[h]
\caption{Lattice results at leading order (LO) and experimental values for the
root-mean-square charge radii and quadrupole moments of the $^{12}$C
states. \label{radii}}
\vspace{.2cm}
\begin{tabular}{|c||c|c|} \hline
& \hspace{.4cm} LO \hspace{.4cm} 
& \hspace{.4cm} Exp \hspace{.4cm} \\ \hline
$r(0_{1}^{+})$ [fm] & $2.2(2)$ & $2.47(2)$ \cite{Schaller:1982} \\
$r(2_{1}^{+})$ [fm] & $2.2(2)$ & $-$ \\
$Q(2_{1}^{+})$ [$e\,$fm$^{2}$] & $6(2)$ & $6(3)$ \cite{Vermeer:1983}\\\hline
$r(0_{2}^{+})$ [fm] & $2.4(2)$ & $-$ \\
$r(2_{2}^{+})$ [fm] & $2.4(2)$ & $-$ \\
$Q(2_{2}^{+})$ [$e\,$fm$^{2}$] & $-7(2)$ & $-$ \\ \hline
\end{tabular}
\end{table}

The LO results for the electromagnetic transitions among the
even-parity states of $^{12}$C are shown in Table~\ref{transitions}. For a
definition of the quantities shown, see {\it e.g.} Ref.~\cite{BohrMottelson:1969v1}.
We find reasonable agreement with available experimental data. The lattice 
results at LO have a tendency to slightly underestimate the experimental values.
Presumably, this reflects the greater binding energies and smaller radii of the 
nuclei at LO. We also predict electromagnetic decays involving the $2_{2}^{+}$ state, 
which are likely to be measured experimentally in the near future.
\begin{table}[h]
\caption{Lattice results at leading order (LO) and experimental values for
electromagnetic transitions involving the even-parity states of $^{12}$C. 
\label{transitions}}
\vspace{.2cm}
\begin{tabular}{|l||c|c|} \hline
& \hspace{.4cm} LO \hspace{.4cm} 
& \hspace{.4cm} Exp \hspace{.4cm} \\ \hline
$B(\mathrm{E2},2_{1}^{+}\rightarrow0_{1}^{+})$ [$e^{2} $fm$^{4}$] & $5(2)$ & $7.6(4)$
\cite{Ajzenberg:1990a} \\
$B(\mathrm{E2},2_{1}^{+}\rightarrow0_{2}^{+})$ [$e^{2} $fm$^{4}$] & $1.5(7)$ &
$2.6(4)$ \cite{Ajzenberg:1990a} \\ \hline
$B(\mathrm{E2},2_{2}^{+}\rightarrow0_{1}^{+})$ [$e^{2} $fm$^{4}$] & $2(1)$ &
$-$ \\
$B(\mathrm{E2},2_{2}^{+}\rightarrow0_{2}^{+})$ [$e^{2} $fm$^{4}$] & $6(2)$ &
$-$ \\ \hline
$m(\mathrm{E0},0_{2}^{+}\rightarrow0_{1}^{+})$ [$e\,$fm$^{2}$] & $3(1)$ & $5.5(1)$
\cite{Chernykh:2010zu} \\ \hline
\end{tabular}
\end{table}

In summary, we have presented \textit{ab initio}~lattice calculations which
reveal the structure of the Hoyle state and find evidence for a low-lying spin-2
rotational excitation. For the ground state and the first spin-2 state, we find
mostly a compact triangular configuration of alpha clusters. For the Hoyle
state and the second spin-2 state, we find a ``bent-arm'' or obtuse triangular
configuration of alpha clusters. We have calculated charge radii, quadrupole
moments and electromagnetic transitions among the low-lying even-parity
states of $^{12}$C at LO. All of our results are in reasonable
agreement with experiment. While further work is clearly needed (such as
calculations using smaller lattice spacings), our results provide a
deeper understanding starting from first principles, of the structure and rotations 
of the Hoyle state.

\begin{acknowledgments}
We thank W.~Nazarewicz, T.~Neff, G.~Rupak, 
H.~Weller, and W.~Zimmerman for useful discussions. Partial financial support
from the DFG and NSFC (CRC~110), HGF (VH-VI-417), BMBF (06BN7008) USDOE
(DE-FG02-03ER41260), EU HadronPhysics3 project \textquotedblleft Study of
strongly interacting matter\textquotedblright, and ERC project 259218
NUCLEAREFT. Computational resources were provided by the J\"{u}lich
Supercomputing Centre (JSC) at the Forschungszentrum J\"{u}lich. 
\end{acknowledgments}


\bibliographystyle{apsrev}
\bibliography{References}

\begin{thebibliography}{29}
\expandafter\ifx\csname natexlab\endcsname\relax\def\natexlab#1{#1}\fi
\expandafter\ifx\csname bibnamefont\endcsname\relax
  \def\bibnamefont#1{#1}\fi
\expandafter\ifx\csname bibfnamefont\endcsname\relax
  \def\bibfnamefont#1{#1}\fi
\expandafter\ifx\csname citenamefont\endcsname\relax
  \def\citenamefont#1{#1}\fi
\expandafter\ifx\csname url\endcsname\relax
  \def\url#1{\texttt{#1}}\fi
\expandafter\ifx\csname urlprefix\endcsname\relax\def\urlprefix{URL }\fi
\providecommand{\bibinfo}[2]{#2}
\providecommand{\eprint}[2][]{\url{#2}}

\bibitem[{\citenamefont{{\"O}pik}(1951)}]{Opik:1951}
\bibinfo{author}{\bibfnamefont{E.~J.} \bibnamefont{{\"O}pik}},
  \bibinfo{journal}{Proc. Roy. Irish Acad.} \textbf{\bibinfo{volume}{A54}},
  \bibinfo{pages}{49} (\bibinfo{year}{1951}).

\bibitem[{\citenamefont{Salpeter}(1952)}]{Salpeter:1952}
\bibinfo{author}{\bibfnamefont{E.~E.} \bibnamefont{Salpeter}},
  \bibinfo{journal}{Astrophys. J.} \textbf{\bibinfo{volume}{115}},
  \bibinfo{pages}{326} (\bibinfo{year}{1952}).

\bibitem[{\citenamefont{Salpeter}(1953)}]{Salpeter:1953}
\bibinfo{author}{\bibfnamefont{E.~E.} \bibnamefont{Salpeter}},
  \bibinfo{journal}{Ann. Rev. Nucl. Sci.} \textbf{\bibinfo{volume}{2}},
  \bibinfo{pages}{41} (\bibinfo{year}{1953}).

\bibitem[{\citenamefont{Hoyle}(1954)}]{Hoyle:1954zz}
\bibinfo{author}{\bibfnamefont{F.}~\bibnamefont{Hoyle}},
  \bibinfo{journal}{Astrophys. J. Suppl.} \textbf{\bibinfo{volume}{1}},
  \bibinfo{pages}{121} (\bibinfo{year}{1954}).

\bibitem[{\citenamefont{Cook et~al.}(1957)\citenamefont{Cook, Fowler,
  Lauritsen, and Lauritsen}}]{Cook:1957}
\bibinfo{author}{\bibfnamefont{C.}~\bibnamefont{Cook}},
  \bibinfo{author}{\bibfnamefont{W.~A.} \bibnamefont{Fowler}},
  \bibinfo{author}{\bibfnamefont{C.~C.} \bibnamefont{Lauritsen}},
  \bibnamefont{and}
  \bibinfo{author}{\bibfnamefont{T.}~\bibnamefont{Lauritsen}},
  \bibinfo{journal}{Phys. Rev.} \textbf{\bibinfo{volume}{107}},
  \bibinfo{pages}{508} (\bibinfo{year}{1957}).

\bibitem[{\citenamefont{Morinaga}(1956)}]{Morinaga:1956}
\bibinfo{author}{\bibfnamefont{H.}~\bibnamefont{Morinaga}},
  \bibinfo{journal}{Phys. Rev.} \textbf{\bibinfo{volume}{101}},
  \bibinfo{pages}{254} (\bibinfo{year}{1956}).

\bibitem[{\citenamefont{Tohsaki et~al.}(2001)\citenamefont{Tohsaki, Horiuchi,
  Schuck, and Ropke}}]{Tohsaki:2001an}
\bibinfo{author}{\bibfnamefont{A.}~\bibnamefont{Tohsaki}},
  \bibinfo{author}{\bibfnamefont{H.}~\bibnamefont{Horiuchi}},
  \bibinfo{author}{\bibfnamefont{P.}~\bibnamefont{Schuck}}, \bibnamefont{and}
  \bibinfo{author}{\bibfnamefont{G.}~\bibnamefont{Ropke}},
  \bibinfo{journal}{Phys. Rev. Lett.} \textbf{\bibinfo{volume}{87}},
  \bibinfo{pages}{192501} (\bibinfo{year}{2001}).

\bibitem[{\citenamefont{Freer et~al.}(2009)}]{Freer:2009}
\bibinfo{author}{\bibfnamefont{M.}~\bibnamefont{Freer}} \bibnamefont{et~al.},
  \bibinfo{journal}{Phys. Rev. C} \textbf{\bibinfo{volume}{80}},
  \bibinfo{pages}{041303} (\bibinfo{year}{2009}).

\bibitem[{\citenamefont{Hyldegaard et~al.}(2010)}]{Hyldegaard:2010}
\bibinfo{author}{\bibfnamefont{S.}~\bibnamefont{Hyldegaard}}
  \bibnamefont{et~al.}, \bibinfo{journal}{Phys. Rev. C}
  \textbf{\bibinfo{volume}{81}}, \bibinfo{pages}{024303}
  (\bibinfo{year}{2010}).

\bibitem[{\citenamefont{Zimmerman et~al.}(2011)\citenamefont{Zimmerman,
  Destefano, Freer, Gai, and Smit}}]{Zimmerman:2011}
\bibinfo{author}{\bibfnamefont{W.~R.} \bibnamefont{Zimmerman}},
  \bibinfo{author}{\bibfnamefont{N.~E.} \bibnamefont{Destefano}},
  \bibinfo{author}{\bibfnamefont{M.}~\bibnamefont{Freer}},
  \bibinfo{author}{\bibfnamefont{M.}~\bibnamefont{Gai}}, \bibnamefont{and}
  \bibinfo{author}{\bibfnamefont{F.~D.} \bibnamefont{Smit}},
  \bibinfo{journal}{Phys. Rev. C} \textbf{\bibinfo{volume}{84}},
  \bibinfo{pages}{027304} (\bibinfo{year}{2011}).

\bibitem[{\citenamefont{Itoh et~al.}(2011)}]{Itoh:2011}
\bibinfo{author}{\bibfnamefont{M.}~\bibnamefont{Itoh}} \bibnamefont{et~al.},
  \bibinfo{journal}{Phys. Rev. C} \textbf{\bibinfo{volume}{84}},
  \bibinfo{pages}{054308} (\bibinfo{year}{2011}).

\bibitem[{\citenamefont{Smit et~al.}(2012)\citenamefont{Smit, Nemulodi,
  Buthelezi, Carter, Fearick et~al.}}]{Smit:2012zk}
\bibinfo{author}{\bibfnamefont{F.}~\bibnamefont{Smit}},
  \bibinfo{author}{\bibfnamefont{F.}~\bibnamefont{Nemulodi}},
  \bibinfo{author}{\bibfnamefont{Z.}~\bibnamefont{Buthelezi}},
  \bibinfo{author}{\bibfnamefont{J.}~\bibnamefont{Carter}},
  \bibinfo{author}{\bibfnamefont{R.}~\bibnamefont{Fearick}},
  \bibnamefont{et~al.} (\bibinfo{year}{2012}).

\bibitem[{\citenamefont{Epelbaum et~al.}(2011)\citenamefont{Epelbaum, Krebs,
  Lee, and Mei{\ss}ner}}]{Epelbaum:2011md}
\bibinfo{author}{\bibfnamefont{E.}~\bibnamefont{Epelbaum}},
  \bibinfo{author}{\bibfnamefont{H.}~\bibnamefont{Krebs}},
  \bibinfo{author}{\bibfnamefont{D.}~\bibnamefont{Lee}}, \bibnamefont{and}
  \bibinfo{author}{\bibfnamefont{U.-G.} \bibnamefont{Mei{\ss}ner}},
  \bibinfo{journal}{Phys. Rev. Lett.} \textbf{\bibinfo{volume}{106}},
  \bibinfo{pages}{192501} (\bibinfo{year}{2011}).

\bibitem[{\citenamefont{Roth et~al.}(2011)\citenamefont{Roth, Langhammer,
  Calci, Binder, and Navratil}}]{Roth:2011ar}
\bibinfo{author}{\bibfnamefont{R.}~\bibnamefont{Roth}},
  \bibinfo{author}{\bibfnamefont{J.}~\bibnamefont{Langhammer}},
  \bibinfo{author}{\bibfnamefont{A.}~\bibnamefont{Calci}},
  \bibinfo{author}{\bibfnamefont{S.}~\bibnamefont{Binder}}, \bibnamefont{and}
  \bibinfo{author}{\bibfnamefont{P.}~\bibnamefont{Navratil}},
  \bibinfo{journal}{Phys. Rev. Lett.} \textbf{\bibinfo{volume}{107}},
  \bibinfo{pages}{072501} (\bibinfo{year}{2011}).

\bibitem[{\citenamefont{Forssen et~al.}(2011)\citenamefont{Forssen, Roth, and
  Navratil}}]{Forssen:2011dr}
\bibinfo{author}{\bibfnamefont{C.}~\bibnamefont{Forssen}},
  \bibinfo{author}{\bibfnamefont{R.}~\bibnamefont{Roth}}, \bibnamefont{and}
  \bibinfo{author}{\bibfnamefont{P.}~\bibnamefont{Navratil}}
  (\bibinfo{year}{2011}), \eprint{1110.0634}.

\bibitem[{\citenamefont{Chernykh et~al.}(2007)\citenamefont{Chernykh,
  Feldmeier, Neff, von Neumann-Cosel, and Richter}}]{Chernykh:2007a}
\bibinfo{author}{\bibfnamefont{M.}~\bibnamefont{Chernykh}},
  \bibinfo{author}{\bibfnamefont{H.}~\bibnamefont{Feldmeier}},
  \bibinfo{author}{\bibfnamefont{T.}~\bibnamefont{Neff}},
  \bibinfo{author}{\bibfnamefont{P.}~\bibnamefont{von Neumann-Cosel}},
  \bibnamefont{and} \bibinfo{author}{\bibfnamefont{A.}~\bibnamefont{Richter}},
  \bibinfo{journal}{Phys. Rev. Lett.} \textbf{\bibinfo{volume}{98}},
  \bibinfo{pages}{032501} (\bibinfo{year}{2007}).

\bibitem[{\citenamefont{Chernykh et~al.}(2010)\citenamefont{Chernykh,
  Feldmeier, Neff, von Neumann-Cosel, and Richter}}]{Chernykh:2010zu}
\bibinfo{author}{\bibfnamefont{M.}~\bibnamefont{Chernykh}},
  \bibinfo{author}{\bibfnamefont{H.}~\bibnamefont{Feldmeier}},
  \bibinfo{author}{\bibfnamefont{T.}~\bibnamefont{Neff}},
  \bibinfo{author}{\bibfnamefont{P.}~\bibnamefont{von Neumann-Cosel}},
  \bibnamefont{and} \bibinfo{author}{\bibfnamefont{A.}~\bibnamefont{Richter}},
  \bibinfo{journal}{Phys. Rev. Lett.} \textbf{\bibinfo{volume}{105}},
  \bibinfo{pages}{022501} (\bibinfo{year}{2010}).

\bibitem[{\citenamefont{Epelbaum
  et~al.}(2009{\natexlab{a}})\citenamefont{Epelbaum, Hammer, and
  Mei{\ss}ner}}]{Epelbaum:2008ga}
\bibinfo{author}{\bibfnamefont{E.}~\bibnamefont{Epelbaum}},
  \bibinfo{author}{\bibfnamefont{H.-W.} \bibnamefont{Hammer}},
  \bibnamefont{and} \bibinfo{author}{\bibfnamefont{U.-G.}
  \bibnamefont{Mei{\ss}ner}}, \bibinfo{journal}{Rev. Mod. Phys.}
  \textbf{\bibinfo{volume}{81}}, \bibinfo{pages}{1773}
  (\bibinfo{year}{2009}{\natexlab{a}}).

\bibitem[{\citenamefont{M{\"u}ller et~al.}(2000)\citenamefont{M{\"u}ller,
  Koonin, Seki, and van Kolck}}]{Muller:1999cp}
\bibinfo{author}{\bibfnamefont{H.~M.} \bibnamefont{M{\"u}ller}},
  \bibinfo{author}{\bibfnamefont{S.~E.} \bibnamefont{Koonin}},
  \bibinfo{author}{\bibfnamefont{R.}~\bibnamefont{Seki}}, \bibnamefont{and}
  \bibinfo{author}{\bibfnamefont{U.}~\bibnamefont{van Kolck}},
  \bibinfo{journal}{Phys. Rev.} \textbf{\bibinfo{volume}{C61}},
  \bibinfo{pages}{044320} (\bibinfo{year}{2000}).

\bibitem[{\citenamefont{Lee et~al.}(2004)\citenamefont{Lee, Borasoy, and
  Sch{\"a}fer}}]{Lee:2004si}
\bibinfo{author}{\bibfnamefont{D.}~\bibnamefont{Lee}},
  \bibinfo{author}{\bibfnamefont{B.}~\bibnamefont{Borasoy}}, \bibnamefont{and}
  \bibinfo{author}{\bibfnamefont{T.}~\bibnamefont{Sch{\"a}fer}},
  \bibinfo{journal}{Phys. Rev.} \textbf{\bibinfo{volume}{C70}},
  \bibinfo{pages}{014007} (\bibinfo{year}{2004}).

\bibitem[{\citenamefont{Epelbaum
  et~al.}(2009{\natexlab{b}})\citenamefont{Epelbaum, Krebs, Lee, and
  Mei{\ss}ner}}]{Epelbaum:2009zs}
\bibinfo{author}{\bibfnamefont{E.}~\bibnamefont{Epelbaum}},
  \bibinfo{author}{\bibfnamefont{H.}~\bibnamefont{Krebs}},
  \bibinfo{author}{\bibfnamefont{D.}~\bibnamefont{Lee}}, \bibnamefont{and}
  \bibinfo{author}{\bibfnamefont{U.~G.} \bibnamefont{Mei{\ss}ner}},
  \bibinfo{journal}{Eur. Phys. J.} \textbf{\bibinfo{volume}{A41}},
  \bibinfo{pages}{125} (\bibinfo{year}{2009}{\natexlab{b}}).

\bibitem[{\citenamefont{Epelbaum
  et~al.}(2010{\natexlab{a}})\citenamefont{Epelbaum, Krebs, Lee, and
  Mei{\ss}ner}}]{Epelbaum:2009pd}
\bibinfo{author}{\bibfnamefont{E.}~\bibnamefont{Epelbaum}},
  \bibinfo{author}{\bibfnamefont{H.}~\bibnamefont{Krebs}},
  \bibinfo{author}{\bibfnamefont{D.}~\bibnamefont{Lee}}, \bibnamefont{and}
  \bibinfo{author}{\bibfnamefont{U.-G.} \bibnamefont{Mei{\ss}ner}},
  \bibinfo{journal}{Phys. Rev. Lett.} \textbf{\bibinfo{volume}{104}},
  \bibinfo{pages}{142501} (\bibinfo{year}{2010}{\natexlab{a}}).

\bibitem[{\citenamefont{Lee}(2009)}]{Lee:2008fa}
\bibinfo{author}{\bibfnamefont{D.}~\bibnamefont{Lee}}, \bibinfo{journal}{Prog.
  Part. Nucl. Phys.} \textbf{\bibinfo{volume}{63}}, \bibinfo{pages}{117}
  (\bibinfo{year}{2009}).

\bibitem[{\citenamefont{Wigner}(1937)}]{Wigner:1937}
\bibinfo{author}{\bibfnamefont{E.}~\bibnamefont{Wigner}},
  \bibinfo{journal}{Phys. Rev.} \textbf{\bibinfo{volume}{51}},
  \bibinfo{pages}{106} (\bibinfo{year}{1937}).

\bibitem[{\citenamefont{Epelbaum
  et~al.}(2010{\natexlab{b}})\citenamefont{Epelbaum, Krebs, Lee, and
  Mei{\ss}ner}}]{Epelbaum:2010xt}
\bibinfo{author}{\bibfnamefont{E.}~\bibnamefont{Epelbaum}},
  \bibinfo{author}{\bibfnamefont{H.}~\bibnamefont{Krebs}},
  \bibinfo{author}{\bibfnamefont{D.}~\bibnamefont{Lee}}, \bibnamefont{and}
  \bibinfo{author}{\bibfnamefont{U.-G.} \bibnamefont{Mei{\ss}ner}},
  \bibinfo{journal}{Eur. Phys. J.} \textbf{\bibinfo{volume}{A45}},
  \bibinfo{pages}{335} (\bibinfo{year}{2010}{\natexlab{b}}).

\bibitem[{\citenamefont{Schaller et~al.}(1982)\citenamefont{Schaller,
  Schellenberg, Phan, Piller, Ruetschi, and Schneuwly}}]{Schaller:1982}
\bibinfo{author}{\bibfnamefont{L.~A.} \bibnamefont{Schaller}},
  \bibinfo{author}{\bibfnamefont{L.}~\bibnamefont{Schellenberg}},
  \bibinfo{author}{\bibfnamefont{T.~Q.} \bibnamefont{Phan}},
  \bibinfo{author}{\bibfnamefont{G.}~\bibnamefont{Piller}},
  \bibinfo{author}{\bibfnamefont{A.}~\bibnamefont{Ruetschi}}, \bibnamefont{and}
  \bibinfo{author}{\bibfnamefont{H.}~\bibnamefont{Schneuwly}},
  \bibinfo{journal}{Nucl. Phys.} \textbf{\bibinfo{volume}{A379}},
  \bibinfo{pages}{523} (\bibinfo{year}{1982}).

\bibitem[{\citenamefont{Vermeer et~al.}(1983)}]{Vermeer:1983}
\bibinfo{author}{\bibfnamefont{W.~J.} \bibnamefont{Vermeer}}
  \bibnamefont{et~al.}, \bibinfo{journal}{Phys. Lett.}
  \textbf{\bibinfo{volume}{B122}}, \bibinfo{pages}{23} (\bibinfo{year}{1983}).

\bibitem[{\citenamefont{Bohr and Mottelson}(1969)}]{BohrMottelson:1969v1}
\bibinfo{author}{\bibfnamefont{A.}~\bibnamefont{Bohr}} \bibnamefont{and}
  \bibinfo{author}{\bibfnamefont{B.~R.} \bibnamefont{Mottelson}},
  \emph{\bibinfo{title}{Nuclear Structure. Volume I: Single-Particle Motion}}
  (\bibinfo{publisher}{W. A. Benjamin}, \bibinfo{address}{New York},
  \bibinfo{year}{1969}).

\bibitem[{\citenamefont{Ajzenberg-Selove}(1990)}]{Ajzenberg:1990a}
\bibinfo{author}{\bibfnamefont{F.}~\bibnamefont{Ajzenberg-Selove}},
  \bibinfo{journal}{Nucl. Phys.} \textbf{\bibinfo{volume}{B506}},
  \bibinfo{pages}{1} (\bibinfo{year}{1990}).

\end{thebibliography}

\end{document}